\documentclass[12pt,thmsa]{article}
\usepackage{amssymb}

\begin{document}

\title{Finite Energy Superluminal Solutions of Maxwell Equations}
\author{E. Capelas de Oliveira$^{1}\thanks{%
e-mail: capelas@ime.unicamp.br}$ \hspace{0.01in}\hspace{0.01in}and W. A.
Rodrigues, Jr.$^{2}$\thanks{%
e-mail: W.Rodrigues@liverpool.ac.uk }\thanks{%
On leave in absence: Institute of Mathematics, Statistics and Scientific
Computation,\emph{\ UNICAMP }and 
Wernher von Braun Advanced Research Center, \emph{UNISAL.}} \\
$^{1}$ Institute of Mathematics, Statistics and Scientific Computation,\\
\emph{IMECC-UNICAMP}\\
CP 6065, 13083-970, Campinas, SP, Brazil\\
$\hspace{-1.2in}^{2}$Department of Mathematics, University of Liverpool\\
Liverpool L69 3BX, UK}
\maketitle

\begin{abstract}
We exhibit exact finite energy superluminal solutions of Maxwell equations
in vacuum and discuss the physical meaning of these solutions.
\end{abstract}

Recently, some papers [1,2] have appeared in the literature showing that in
some hypothetical media there is the possibility of the existence of
superluminal electromagnetic pulses (solutions of Maxwell equations) such
their \emph{fronts} travel in the media with \emph{superluminal} velocities.
Now, the solutions discovered in [1,2], despite their theoretical interest
have \emph{infinite} energy and as such cannot be produced in the physical
world. Only finite aperture approximations to these waves can eventually be
produced (supposing the existence of the special media). The objective of
this letter is to show that in contrast to the solutions discovered in [1,2]
(that, as already said have infinite energy), there exist vacuum solutions
of Maxwell equations which are \emph{finite energy superluminal solutions} .
These new solutions, as we shall see, appear when we solve some Sommerfeld
like problems$^{3,4}$ to be reported below. We discuss if the new solutions
can be realized in the physical world. Moreover, we emphasize that the new
solutions correspond to phenomenon distinct to already observed wave motion
with superluminal [5-8] (or even negative [9,10]) group velocities. In the
case,e.g., of experiments [5-8] only the peaks of the waves travel (for a
while) with superluminal velocity whereas their fronts always travel at the
velocity of light.

We start by recalling how to write electromagnetic field configurations in
terms of Hertz potentials [11,12]. Suppose we have a Hertz potential $\vec{%
\Pi}_{m}$ of magnetic type. In what follows we use units such that the
velocity of light is $c=1$. Then, the associated electromagnetic field is
given by 
\begin{equation}
\vec{E}=-\frac{\partial }{\partial t}(\nabla \times \vec{\Pi}_{m}),\hspace{%
0.2in}\vec{B}=\nabla \times \nabla \times \vec{\Pi}_{m}.  \label{1}
\end{equation}
Let us take $\vec{\Pi}_{m}=\Phi \hat{e}_{z}$. Then, since the Hertz
potential (in vacuum) satisfies a homogeneous wave equation, we have that 
\begin{equation}
\square \Phi =0.  \label{2}
\end{equation}

The \emph{Sommerfeld} problem (not to be confused with a \emph{Cauchy}
problem) to be considered here is the following. In a given inertial frame
(the laboratory\footnote{%
The laboratory is modeled by time like vector field $L=\frac{\partial }{%
\partial t}\in \sec TM$.}) find a solution $\Phi _{X}:(t,\vec{x})\mapsto C$
(where $C$ is the field of complex numbers) for eq.(2) satisfying the
following boundary conditions\footnote{%
The necessity for these boundary conditions is proved in [13].} at the $z=0$
plane, 
\begin{equation}
\left\{ 
\begin{array}{l}
\Phi _{X}(t,\rho ,0)=\mathbf{T}(t)\int\limits_{-\infty }^{\infty }d\omega
B(\omega )J_{0}(\omega \rho \sin \eta )e^{-i\omega t}, \\ 
\left. \frac{\partial \Phi _{X}(t,\rho ,z)}{\partial z}\right| _{z=0}=i%
\mathbf{T}(t)\cos \eta \int\limits_{-\infty }^{\infty }d\omega B(\omega
)J_{0}(\omega \rho \sin \eta )k(\omega )e^{-i\omega t},
\end{array}
\right.   \label{3}
\end{equation}
where $\mathbf{T}(t)=\left[ \Theta (t+T)-\Theta (t-T)\right] $ , $\Theta $
is the Heaviside function, $k(\omega )=\omega $, and $\eta $ is a constant
called the axicon angle [13-18] and $B(k)$ is an appropriate frequency
distribution. As showed in [14] the solution of eq.(2) (for $z>0,t>T$) which
satisfies the Sommerfeld conditions is

\begin{equation}
\Phi _{X}(t,\rho ,z)=\left\{ 
\begin{array}{l}
\int\limits_{-\infty }^{\infty }d\omega B(\omega )J_{0}(\omega \rho \sin
\eta )\emph{e}^{-i\omega (t-z\cos \eta )}\hspace{0.1in}\mbox{for\hspace{0.1in%
}}\left| t-z\cos \eta \right| <T \\ 
\hspace{1in}0\hspace{0.1in}\hspace{0.08in}\hspace{1in}\hspace{0.1in}\mbox{for%
\hspace{0.1in}}\left| t-z\cos \eta \right| >T.
\end{array}
\right.  \label{4}
\end{equation}

We call $\Phi _{X}$ a scalar superluminal $X$-pulse. Now, as is well known,
the energy density for a complex field configuration, like the $\Phi _{X}$,
is 
\begin{equation}
u=(\partial _{t}\Phi _{X})(\partial _{t}\Phi _{X}^{*})+(\partial _{x}\Phi
_{X})(\partial _{x}\Phi _{X}^{*})+(\partial _{y}\Phi _{X})(\partial _{y}\Phi
_{X}^{*})+(\partial _{z}\Phi _{X})(\partial _{z}\Phi _{X}^{*}),  \label{5}
\end{equation}
and the energy of the field configuration can be calculated by the volume
integral of $u$ on a constant time hyperplane, say $t=T^{\prime }>T$. The
calculation is easy when done in \emph{cylindrical} coordinates. Recalling
that from eq.(4) it follows that the support of the pulse at $t=T^{\prime }$
is $\triangle z=2T/\cos \eta $, we have 
\begin{equation}
\mathcal{E}=\frac{8\pi T}{\sin ^{2}\eta \cos \eta }\int\limits_{-\infty
}^{\infty }\left| B(k)\right| ^{2}kdk,  \label{6}
\end{equation}
where the kinetic and potential energy terms give equal contributions.
Eq.(6) gives \emph{finite} energy for the scalar $X$-pulse for an infinity
of frequency distribution functions $B(k)$, such that $\left| B(k)\right|
^{2}$ be null for $k<0$. A trivial example is $B(k)=[$ $\Theta (k)-\Theta
(k-k_{0})]$, with $k_{0}$ a constant. Now, we study the electromagnetic
case. The non null components of the electromagnetic field\footnote{%
Called a superluminal electromagnetic $X$ pulse [13,14].} corresponding to a
magnetic Hertz potential $\vec{\Pi}_{m}=\Phi _{X}\hat{e}_{z}$ are (for $z>0$%
, $t>T)$%
\begin{equation}
\left\{ 
\begin{array}{ll}
\begin{array}{l}
E_{\theta }=i\sin \eta \int\limits_{-\infty }^{\infty
}dkB(k)k^{2}J_{1}(k\rho \sin \eta )e^{-ik(t-z\cos \eta )}, \\ 
B_{\rho }=\frac{-i}{2}\sin 2\eta \int\limits_{-\infty }^{\infty
}dkB(k)k^{2}J_{1}(k\rho \sin \eta )e^{-ik(t-z\cos \eta )}, \\ 
B_{z}=\sin ^{2}\eta \int\limits_{-\infty }^{\infty }dkB(k)k^{2}J_{0}(k\rho
\sin \eta )e^{-ik(t-z\cos \eta )},
\end{array}
& \mbox{for }|t-z\cos \eta |<T, \\ 
E_{\theta }=B_{\rho }=B_{z}=0, & \mbox{for }|t-z\cos \eta |>T.
\end{array}
\right.  \label{7}
\end{equation}

Now, using the standard energy density of the electromagnetic field [11,12],
the energy of the superluminal electromagnetic $X$ pulse results, 
\begin{eqnarray}
\mathcal{E}_{X} &=&\frac{1}{2}\int\limits_{0}^{2\pi }\int\limits_{z_{\min
}}^{z_{\max }}\int\limits_{0}^{\infty }\left[ E_{\theta }E_{\theta
}^{*}+B_{\rho }B_{\rho }^{*}+B_{z}B_{z}^{*}\right] \rho d\rho dzd\theta 
\nonumber \\
&=&\frac{4\pi T}{\cos \eta }\int\limits_{-\infty }^{\infty }\left|
B(k)\right| ^{2}k^{3}dk.  \label{8}
\end{eqnarray}

Eq.(8) gives finite energy for superluminal solutions of Maxwell equations
satisfying Sommerfeld boundary conditions (here expressed through conditions
for the associated Hertz potential) for an infinity of possible frequency
distributions $B(k)$, as in the scalar case.

We have four comments before ending this letter:

(i) What does our finite energy solution (for the scalar wave equation) look
like for an observer in a Lorentz frame $Z\in \sec TM,$%
\begin{equation}
Z=\frac{1}{\sqrt{1-V^{2}}}(\partial _{t}+V\partial _{z}),  \label{9}
\end{equation}

which is moving with velocity $V=\cos \eta $ relative to the laboratory (the
frame $L=\partial _{t}\in \sec TM$ )?

As can be easily verified the transformed solution is: 
\begin{equation}
\Phi _{X}^{\prime }(t^{\prime },\rho ,z^{\prime })=\left\{ 
\begin{array}{l}
\int\limits_{-\infty }^{\infty }d\omega B(\omega )J_{0}(\omega \rho \sin
\eta )\emph{e}^{-i\omega \sin \eta t\acute{}}\hspace{0.1in}\mbox{for\hspace{%
0.1in}}\left| t^{\prime }\right| <T/\sin \eta  \\ 
\hspace{1in}0\hspace{0.1in}\hspace{0.08in}\hspace{1in}\hspace{0.1in}\mbox{for%
\hspace{0.1in}}\left| t^{\prime }\right| >T/\sin \eta .
\end{array}
\right.   \label{10}
\end{equation}

The solution is independent of the spatial coordinate $z$ and corresponds to
a standing wave occupying all the rest space of the $Z$ frame and that
exists only for the time interval $\triangle t^{\prime }=2T/\sin \eta $. Is
this result  non physical? If not, what is the meaning of such a wave for
the observers of the $Z$ frame? As a Minkowski diagram can show, the wave
stands for a finite period of time according to the time order of the $Z$
frame because it is going to the \emph{past} of the $Z$'s observers. This
must be a normal phenomenon if relativity theory is true and \emph{genuine}
superluminal motion exists. The observers at the $Z$ frame will compute an
infinite energy for that wave, but since they \emph{know} relativity theory
they will interpret the whole phenomena as follows: the wave that stands for
a finite period of time at our frame is a superluminal finite energy wave
produced in a laboratory ( the $L$ frame) that is moving with velocity $%
-1/\cos \eta $ relative to our frame (i.e., $Z$ frame). Of course, the $Z$
frames physicists cannot produce such a wave in their frame, due to two
reasons. The first is that the wave according to them has infinite energy
and the second, which is the crucial one, is simply because the device which
produced it is at rest in another frame (the $L$ frame). According to the
Principle of Relativity the $Z$ frame physicists can duplicate in their
frame the device used in the $L$ frame and launch a wave like the one given
by eq.(\ref{4}) (with boundary conditions like in eq.(\ref{3})) with the $%
(t,\rho ,z)$ substituted by $(t^{\prime },\rho ,z^{\prime })$. Of course, if
that would be possible, we would arrive at well known paradoxical situations%
\footnote{%
More details on this issue can be found in [18].}, that fortunately need not
to be discussed here (see (iii) below).

Note also that the $Z$ frame mathematicians aware of the interpretation
given by their fellow physicists can obtain directly the solution given by
eq.(\ref{10}) by solving a generalized \emph{mixed} boundary value problem,
where the boundary conditions are: 
\begin{eqnarray}
&&\left. \Phi _{X}^{\prime }(t^{\prime },\rho ,z^{\prime })\right|
_{z^{\prime }=-\cos \eta t^{\prime }}  \nonumber \\
&=&\left[ \Theta (\sin \eta t^{\prime }+T)-\Theta (\sin \eta t^{\prime
}-T)\right] \int\limits_{-\infty }^{\infty }d\omega B(\omega )J_{0}(\omega
\rho \sin \eta )e^{-i\omega \sin \eta t^{\prime }}  \nonumber \\
&&\left( \gamma \frac{\partial }{\partial z^{\prime }}-\gamma V\frac{%
\partial }{\partial t^{\prime }}\right) \left. \Phi _{X}^{\prime }(t^{\prime
},\rho ,z^{\prime })\right| _{z^{\prime }=-\cos \eta t^{\prime }}  \nonumber
\\
&=&i\cos \eta \left[ \Theta (\sin \eta t^{\prime }+T)-\Theta (\sin \eta
t^{\prime }-T)\right] \int\limits_{-\infty }^{\infty }d\omega B(\omega
)\omega J_{0}(\omega \rho \sin \eta )e^{-i\omega \sin \eta t^{\prime }}. 
\nonumber 
\end{eqnarray}

(ii) Of course, an analogous analysis holds for the finite energy
superluminal solutions of Maxwell equations that we have just found. It is
worth saying here that the existence of such solutions does not conflict
with the famous result on the Cauchy problem concerning the Maxwell
equations. That result says: any electromagnetic field configuration with
compact support at $t=0,$ let us say for $\left| \vec{x}\right| \leq R$, is
such that the field is null for $t>0$ for all $\left| \vec{x}\right| \geq
R+t $.\footnote{%
A proof of an anlagous theorem for the homogeneous wave equation can be
found in [13]. For Maxwell equations see [14].}

(iii) Is it possible to build a physical device to launch a finite energy
superluminal electromagnetic $X$ pulse? Our answer is \emph{no}. Indeed,
finite aperture approximations (FAA) to exact superluminal $X$-like
solutions of Maxwell equations (which, of course have finite energy) have
already been produced [7,8]. However, these FAA are such that their peaks
move with velocity $v>1$ but their front always moves with the speed of
light. This result has been predicted in [16,18] and is endorsed by the
experimental results of [7,8] as proved in [13]. Now, concerning the
solutions we just found, in order for them to be produced (by an antenna) as
real physical waves it is necessary to produce waves that extend in all the $%
z=0$ plane where the antenna is located for the time interval $-T<t<T$. Of
course, this is physically impossible because it would require that the
antenna should be an infinite one.

(iv) Besides the superluminal solutions just found, there are also finite
energy \emph{subluminal} solutions (to be reported elsewhere). We must say
that  even if the new superluminal solutions cannot be produced by physical
devices  the only possible reason for their \emph{non} existence   in our
universe is that of a possible violation of the principle of relativity.
Eventually these new superluminal solutions may also find applications in
the understanding of some fundamental issues concerning the nonlocality
problem in quantum mechanics [21].\bigskip 

\noindent{\bf Acknowledgments}: Authors are grateful to Drs. D.S. Thober and A.L.Xavier for discussions and to Dr. I. Porteous for a careful reading of the
manuscript.

\noindent{\bf References}

\begin{enumerate}
\item  P. Ghose and  M.K. Samal, \emph{Lorentz Invariant Superluminal Tunneling}. in publ. \emph{Phys. Rev. E }\textbf{64, \# }036620\emph{\ }%
(2001).(http://www.arXiv:quant-ph/0011033 v3).

\item X. Zhou, Possibility of a light pulse with speed greater than $c$, \emph{%
Phys. Lett. A} \textbf{278, }1-5\textbf{\ }(2001)

\item L. Brillouin, \emph{Wave Propagation and Group Velocity}, Academic
Press, New York, 1960.

\item F.A. Mehmeti,\emph{Transient Tunnel Effect and Sommerfeld Problem},
Akademie Verlag, Berlin, 1996.

\item A. Enders,  G.Nimtz, On superluminal barrier traversal, \emph{J. Phys. I
(France)} \textbf{2}, 1693-1698 (1992).

\item A. M. Steinberg,  P.G. Kwiat, and  R.Y. Chiao, Measurement of the
single-photon tunneling time, \emph{Phys. Rev. Lett.}\textbf{71}, 708-711
(1993)

\item P. Saari and  K. Reivelt, Evidence of $X$-shaped propagation-invariant
localized light waves, \emph{Phys. Rev. Lett}. \textbf{21,} 4135-4138 (1997).

\item D. Mugnai , A. Ranfagni, and R. Ruggeri, Observation of superluminal
behaviors in wave propagation, \emph{Phys. Rev. Lett.} \textbf{80, }%
4830-4833 (2000).

\item E. L. Bolda,  J. C. Garrison, and R.Y. Chiao, Optical pulse propagation
at negative group velocities due to nearby gain line, \emph{Phys.Rev. A }%
\textbf{49}, 2938-2947 (1994)

\item L.J. Wang,  A. Kumzmich, and A. Dogariu, Gain-assisted superluminal
light propagation, \emph{Nature} \textbf{406}, 277-279 (2000).

\item J. A. Stratton,  \emph{Electromagnetic Theory}, McGraw-Hill, New York,
1941.

\item W.K.H. Panofski, and  M. Phillips, \emph{Classical Electricity and
Magnetism}, 2nd ed., Addison-Wesley, Reading, MA, 1962.

\item W.A. Rodrigues Jr., D.S. Thober, and A.L. Xavier Jr., Causal explanation
of superluminal behavior of microwave propagation in free space, \emph{Phys.
Lett. A} \textbf{284}, 217-224 (2001).

\item E. Capelas de Oliveira, W.A. Rodrigues Jr., D.S. Thober and A.L. Xavier
Jr., Thoughtful comments on `Bessel beams and signal propagation', \emph{%
Phys. Lett. A} \textbf{284}, 296-303 (2001).

\item J. Durnin, Exact solutions for nondiffracting beams.1.The scalar theory, 
\emph{J. Opt. Soc. Am. A} \textbf{4}, 651-654 (1987).

\item W.A. Rodrigues Jr., and  J.Y. Lu, On the existence of undistorted
progressive waves (UPWs) of arbitrary speeds $0\leq v<\infty $ in nature. 
\emph{Found. Phys.} \textbf{27,} 435-503 (1997).

\item E. Capelas Oliveira, and W A. Rodrigues Jr., Superluminal
electromagentic waves in free space, \emph{Ann. der Physik} \textbf{7,}
654-659 (1998).

\item J. E.  Maiorino, and  W.A. Rodrigues Jr., \emph{What is Superluminal
Wave Motion}? electronic book at http://www.cptec.br/stm, \emph{Sci. and
Tech. Mag. }\textbf{4}(2) (1999\emph{).}

\item M.E. Taylor, \emph{Pseudo Differential Operators}. Princeton Univ.
Press, Princeton, 1981.

\item R. Courant, and D. Hilbert, \emph{Methods of Mathematical Physics}, vol. 
\textbf{2}. John Wiley and Sons, New York, 1966.

\item  A.A. Grib,  and W.A. Rodrigues Jr., \emph{Nonlocality in Quantum Physics}. Kluwer Acad./Plenum Publ., New York, 1999.
\end{enumerate}
\end{document}